\documentclass[fleqn,twoside]{article}
\usepackage{espcrc2}

\usepackage{graphicx}
\usepackage[figuresright]{rotating}

\title{Hadronic $B$ decays}

\author{M. Beneke\address[]{Institut f\"ur Theoretische Physik E,
    RWTH Aachen, Sommerfeldstr. 28, D - 52074 Aachen, Germany}%
        \thanks{Talk presented at ``Beauty 2006'', 
                September 25th -- 29th 2006, Oxford, U.K. and 
                at  ``Heavy Quarks \&
                Leptons'', October 16th -- 20th 2006, Munich, Germany. }}

\begin{document}

\begin{abstract}
\noindent 
I briefly summarize the factorization approach to hadronic $B$ decays 
emphasizing theoretical results that have become available recently. 
The discussion of its application to data is abridged, and only the 
determination of $\gamma=(71\pm 5)^\circ$ from time-dependent 
CP asymmetries is included in some detail.
\vspace{1pc}
\end{abstract}

\maketitle

\vspace*{-7.2cm}
\noindent PITHA 06/14\\
\phantom{hep-ph/0612353}
%hep-ph/0612353
\vspace*{5.2cm}

\section{INTRODUCTION}
\setcounter{footnote}{0}

\noindent
Many observables at the $B$ factories are connected with 
branching fractions, CP asymmetries and polarization of exclusive, 
hadronic $B$ decays. They provide access to the flavour and spin structure 
of the weak interaction, but a straightforward interpretation is 
usually obscured by the strong interaction. In technical terms, 
the difficult (long-distance) part of the strong interaction 
resides in the matrix elements $\langle f|O_i|\bar B\rangle$, 
where $O_i$ is an operator in the effective weak interaction 
Lagrangian.

Systematic approaches to hadronic $B$ decays are based on expansions 
in small parameters. The two available options exploit approximate flavour 
symmetries (expansion parameter $m_q/\Lambda$, $m_q$ a light 
quark mass), or the large energy transfer in $B$ decays (expansion 
parameter $\Lambda/m_b$), resulting in two frameworks -- 
``SU(3)'' and ``Factorization'' -- that 
could hardly be different methodically and technically. In practice, 
both frameworks are implemented only at the leading order, and additional 
assumptions are usually necessary (neglecting ``small'' amplitudes; 
estimating $\Lambda/m_b$ corrections). Despite this restriction, there 
has been much progress by applying and working out these theories 
over the past few years. In the following I focus on the factorization 
approach. Furthermore, $f$ will be assumed to be a charmless, 
two-body, meson final state; the mesons are assumed to be 
pseudoscalar or vector mesons from the ground state nonet.

\section{THEORY OF HADRONIC DECAYS (FACTORIZATION)}

\noindent 
The starting point is the investigation of Feynman diagrams 
with external collinear lines (energetic, massless lines with momenta 
nearly parallel to one of the two final state mesons, $M_1$ or $M_2$), 
one nearly on-shell heavy-quark line, and soft lines (representing 
the light degrees of freedom in the $\bar B$ meson). The simultaneous 
relevance of collinear and soft configurations implies three 
relevant scales: $m_b$, $\sqrt{m_b \Lambda}$, and $\Lambda$. 
In the heavy-quark limit the first two are perturbative, and only 
the third is long-distance. Factorization amounts to 
showing that the long-distance contributions to the matrix elements 
$\langle [c1][c2]|O_i|[s]\rangle$ are actually contained in 
the simpler matrix elements $\langle [c1]|(\bar q b)(0)|[s]\rangle$ 
(form factors), $\langle [ci]|\bar q(x)q(0)|0\rangle$, 
and  $\langle 0|\bar q(x)b(0)|[s]\rangle$ (light-cone 
distribution amplitudes). It is then assumed that if this holds 
perturbatively to all orders for all quark-gluon matrix elements, 
then it does for the hadronic matrix elements.

Factorization in a similar form was first applied to $B$ decays in 
\cite{Fakirov:1977ta} as a phenomenological approximation akin 
to the vacuum saturation approximation for the four-quark 
operator matrix elements relevant to $B\bar B$ mixing. Intuitively, 
factorization might work, because the partons that eventually 
form the meson $M_2$ that does 
not pick up the spectator quark escape the $\bar B$ remnant 
as an energetic, low-mass, colour-singlet system, and hadronize 
far away and therefore independently from the remnant. This 
qualitative argument was given in \cite{Bjorken:1988kk} for 
the decay $\bar B_d\to D^+\pi^-$. In \cite{BBNS1999}~it was shown 
to hold for charmless decays, where the disruption of the 
$B$ meson is much more violent, and a calculational 
framework was provided, in which the phenomenological factorization 
approach was contained as a leading-order approximation. At the 
same time, the next-to-leading order corrections were computed. 

The new factorization formula included a new mechanism, 
spectator-scattering, where a hard-collinear interaction 
with the soft remnant takes place. Thanks to the development of 
soft-collinear effective theory (SCET), this mechanism is now much 
better understood. In the following I sketch the rederivation of 
the factorization formula in SCET~\cite{Chay:2003ju,BF03,Bauer04}.

Integrating out fluctuations on the scale $m_b$ at leading power 
in the $\Lambda/m_b$ expansion amounts to an analysis of the 
structure of hard subgraphs with external hard-collinear, collinear 
and soft lines. Those identified as leading are then 
calculated perturbatively in $\alpha_s(m_b)$. Formulated as an 
operator matching equation from QCD to SCET${}_{\rm I}$, the result of 
this analysis reads 
\begin{eqnarray}
O_i &=& \big[\bar\chi^{(0)}(t n_-)\chi^{(0)}\big] 
*\Big(C_i^{\rm I}(t) \,\big[\bar\xi h_v\big] 
\nonumber\\
&&+ \, C_i^{\rm II}(t,s) * \big[\bar\xi \!\not\!\!A_\perp(s n_+)h_v\big]
\Big).
\label{step1}
\end{eqnarray}
Remarks: (a) The short-distance coefficient $C_i^{\rm I}$ 
incorporates corrections to naive factorization. The term 
in the second line describes spectator-scattering with its 
own short-distance coefficient  $C_i^{\rm II}$. (b) The 
second line is a leading contribution despite the fact 
that the corresponding operator is suppressed 
in dimensional and SCET${}_{\rm I}$ power counting. This 
follows by extension of the power-counting analysis of 
\cite{BF03}. (c) The meson $M_2$ factorizes already below the 
scale $m_b$ \cite{Bauer04}, since SCET${}_{\rm I}$ does not contain 
interactions between the $\chi^{(0)}$ fields and the 
collinear-1 and soft fields. It follows 
that at leading power in the heavy-quark expansion, the 
strong interaction phases originate from the short-distance 
coefficients $C_i^{\rm I,II}$ at the hard scale. (d) The 
result above must be modified to account for a non-factorizing 
effect when the final state contains an $\eta^{(\prime)}$ meson. 
This effect is explained in \cite{Beneke:2002jn}, but 
appears to have been missed in the SCET rederivation 
of the factorization theorem for mesons with 
flavour-singlet components~\cite{Williamson:2006hb}. Taking 
the hadronic matrix element of (\ref{step1}) gives 
\begin{eqnarray}
\langle M_1 M_2|O_i|\bar B\rangle &=& 
\Phi_{M_2}(u) * \Big(T^{\rm I}(u) \,F^{B M_1}(0) 
\nonumber\\ 
&&\hspace*{-2cm}+ \,C^{\rm II}(\tau,u) * \Xi^{B M_1}(\tau,0)\Big),
\label{me}
\end{eqnarray}
where I have reintroduced the full QCD form factor $F^{B M_1}(0)$ 
resulting in a slight modification of the short-distance 
coefficients. $\Xi^{B M_1}(\tau,0)$ denotes a new, unknown, non-local 
form factor, which depends on the convolution variable $\tau$.

The different implementations of factorization can be distinguished 
broadly by their treatment of the different factors in (\ref{me}). 
In the PQCD approach \cite{Keum:2000wi}~the form factors 
$F^{B M_1}(0)$ and $\Xi^{B M_1}(\tau,0)$ are assumed to be 
short-distance dominated, and claimed to be calculable in 
a generalized factorization framework ($k_\perp$-factorization).  
All four quantities, $T^{\rm I}, C^{\rm II}, F^{B M_1}(0), 
\Xi^{B M_1}(\tau,0)$ have been calculated at leading order. 
Recently, some next-to-leading order (NLO) corrections to 
$T^{\rm I}$ have been included. In the QCD factorization approach 
\cite{BBNS1999} it is assumed that the standard heavy-to-light 
form factors receive a leading soft contribution, and are therefore 
not calculable. However, $\Xi^{B M_1}(\tau,0)$ is dominated by 
perturbative hard-collinear interactions, and factorizes further 
into light-cone distribution amplitudes (see below). In the BBNS 
implementation of QCD factorization, $F^{B M_1}(0)$ is a phenomenological 
input (usually from QCD sum rules). The other three quantities, 
$T^{\rm I}, C^{\rm II}, \Xi^{B M_1}(\tau,0)$ have  been 
calculated at the next-to-leading order. In the BPRS 
implementation~\cite{Bauer04} the use of perturbation theory at the 
hard-collinear scale $\sqrt{m_b\Lambda}$ is avoided, and 
both form factors are fit to hadronic $B$ decay data. This approach is 
restricted to leading-order in the short-distance coefficients, 
since only then does the unknown form factor $\Xi^{B M_1}(\tau,0)$ enter 
the equations through a single moment. There is another difference 
between BBNS and BPRS, who claim that (\ref{step1}) is not valid  
for diagrams with internal charm quark loops. (This should be 
distinguished from \cite{Ciuchini:2001gv}, which speculates about 
large power corrections from charm loops or annihilation.) I believe 
that the theoretical arguments leading to this conclusion are wrong 
\cite{Beneke:2004bn}. For phenomenology, the important consequence 
from treating charm loop diagrams as non-perturbative is that the 
penguin amplitudes must be determined from data, such that no 
CP asymmetry can be predicted from theory alone. Since the tree amplitudes 
are also determined from data (namely, through the two form factors; 
the phase of $C/T$ is automatically zero in a leading-order 
treatment), the BPRS 
approach has much more in common with amplitude fits to data than 
with QCD/SCET calculations. 

The QCD factorization argument is completed by noting that the 
non-local SCET${}_{\rm I}$ form factor $\Xi^{B M_1}(\tau,0)$ factorizes 
into light-cone distribution amplitudes, when the hard-collinear 
scale $\sqrt{m_b\Lambda}$ is integrated out~\cite{BF03}. Inserting 
\begin{equation}
\Xi^{B M_1}(\tau,0) = J(\tau;\omega,v) * \Phi_B(\omega) * 
\Phi_{M_1}(v)
\end{equation}
into (\ref{me}) results in the original QCD factorization formula 
with the additional insight that the spectator-scattering kernel 
$T^{\rm II} = H^{\rm II} * J$ factorizes into a hard and hard-collinear 
kernel. The development of SCET was crucial to identify the 
operators  and precise matching prescriptions 
that make the calculation of higher-order corrections to 
spectator-scattering feasible.

\section{HIGHER-ORDER CALCULATIONS}
\label{calcs}

\noindent
On the calculational side one of the main efforts over the past few 
years has been the calculation of one-loop corrections to 
spectator-scattering, which formally represents a next-to-next-to-leading 
contribution in the QCD factorization approach. This programme 
is now complete. The hard-collinear correction to $J$ has been 
calculated in \cite{Hill:2004if,Kirilin:2005xz,Beneke:2005gs}; 
the hard correction to $H^{\rm II}$ in \cite{Beneke:2005vv,Kivel:2006xc}
for the tree amplitudes and in \cite{BJ2} for the QCD penguin 
and electroweak penguin amplitudes. (An earlier calculation of the 
QCD penguin contribution \cite{Li:2005wx} disagrees with \cite{BJ2}.)
The main results are summarized as follows:
(a) The convolution integrals are convergent, which establishes 
factorization of spectator-scattering at the one-loop order. 
(b) Perturbation theory works for spectator-scattering, 
including perturbation theory at the hard-collinear scale. 
(c) The correction enhances the colour-suppressed tree amplitude, 
and reduces the colour-allowed one. This improves the description of the 
tree-dominated decays to pions and $\rho$ mesons. (d) The correction to the 
colour-allowed QCD penguin amplitude is negligible. Thus there is no 
essential change in the predictions of branching fractions and 
CP asymmetries of penguin-dominated decays. 

The evaluation of the colour-suppressed tree amplitude 
gives~\cite{BJ2} 
\begin{eqnarray}
a_2(\pi\pi) &=& 0.18 - [0.15 + 0.08i]_{\rm NLO} 
   \nonumber\\[0.1cm]
   && \hspace*{-1.5cm} + \,\left[ \frac{r_{\rm sp}}{0.485} \right]
   \Big\{ [0.12]_{\rm LO} + [0.05 +0.05i]_{\rm NLO} 
   \nonumber\\
   && \hspace*{0.4cm} + 
   [0.07]_{\rm tw3} \Big\}. 
\label{a2}
\end{eqnarray}
Here $r_{\rm sp} = (9 f_{M_1}\hat f_B)/(
m_b F^{B M_1}(0) \lambda_B)$ defines a combination of hadronic 
parameters that normalizes the spectator-scattering effect. 
Eq.~(\ref{a2}) shows the importance of computing quantum 
corrections: the naive factorization value 0.18 is nearly 
canceled by the 1-loop vertex correction calculated in 
\cite{BBNS1999}. It now appears that the colour-suppressed 
tree amplitude is generated by spectator-scattering. It is not 
excluded that $r_{\rm sp}$ is a factor of two larger than 
0.485, in which case $a_2$ becomes rather large. My interpretation 
of the pattern of the $\pi\pi$, $\pi\rho$ and $\rho\rho$ 
branching fractions is that spectator-scattering is 
important \cite{BN2003}. On the other hand, the large 
direct CP asymmetry in $\bar B_d\to\pi^+\pi^-$ cannot 
be explained by known radiative corrections, and remains a problem. 

Next-to-leading order corrections have recently been implemented 
in the PQCD approach for the first time~\cite{Li:2005kt}. 
More precisely, the 1-loop kernel $T^{\rm I}$ from the QCD factorization 
approach is used as a short-distance coefficient for the 
subsequent tree-level PQCD calculation. The numerical impact 
is again strongest on the colour-suppressed tree amplitude, $C$. But 
while this correction ($- [0.15 + 0.08i]_{\rm NLO}$ in (\ref{a2})) 
results in a near cancellation of the naive factorization term 
in the QCD factorization approach, it provides an enhancement 
of $C$ by a factor of several in~\cite{Li:2005kt}. This resolves 
the $\pi K$ puzzle in the PQCD approach.

I am rather sceptical about the possibility to perform accurate 
calculations in the PQCD approach. A complete NLO calculation 
in the PQCD approach requires a calculation of all one-loop 
spectator-scattering diagrams (similar 
to~\cite{Beneke:2005vv,Kivel:2006xc,BJ2}) rather than 
the 1-loop BBNS kernels. The calculation of $T^{\rm I}$ is done 
with on-shell external lines, but when the vertex diagram appears as 
a subdiagram in a larger diagram with hard-collinear exchanges, 
the external lines of the subdiagram can be far off-shell. 
Hence $T^I$ is not the appropriate quantity to be used. The 
numerical differences between (\ref{a2}) and~\cite{Li:2005kt} 
despite the same input $T^{\rm I}$ can be traced to the choice 
of scales. The one-loop correction to $T^{\rm I}$ makes the 
result less sensitive to variations of the renormalization scale 
in the Wilson coefficients, but only for scales larger than about 
$1.5\,$GeV, below which perturbation theory breaks down. 
Factorization shows that the scale of the Wilson coefficients 
should be of order $m_b$. However, in the PQCD approach the 
scales $m_b$ and $\sqrt{m_b \Lambda}$ are not distinguished, 
and the Wilson coefficients are evaluated at very low scales 
(to 500 MeV), where perturbation theory is not reliable. An 
unphysical enhancement of the Wilson coefficients at small scales 
is also the origin of the large penguin and annihilation amplitude 
in the PQCD approach. Yet a variation of the renormalization scale 
is not included in theoretical error estimates.

\section{POWER-SUPPRESSED EFFECTS}

\noindent
Power corrections to the QCD penguin amplitudes are essential for 
a successful phenomenology within the factorization framework. The 
most important $\Lambda/m_b$ effect is the scalar QCD penguin 
amplitude $r_\chi a_6$. Fortunately, the bulk contribution to 
this amplitude appears to be calculable, although its factorization 
properties are not yet understood. This power correction is responsible 
for the differences between $PP$, $PV$ and $VV$ final states and the 
$\eta^{(\prime)} K^{(*)}$ final states \cite{Beneke:2002jn}. 
The calculated pattern is 
in very good agreement with experimental data. 

The second most important power correction is presumably weak 
annihilation. I emphasize ``presumably'', since there is no 
unambiguous empirical evidence of any weak annihilation contribution 
in charmless decays, and only upper limits can be derived. The 
theoretical difficulty with power corrections is reflected in the 
different treatments of annihilation. In PQCD it is calculable and 
large. In the BBNS implementation of factorization it is represented 
by a phenomenological parameter \cite{BBNS2001}, not very large, 
but it makes 
the calculation of CP asymmetries uncertain. In the BPRS implementation 
it is neglected together with all power corrections. This is 
phenomenologically viable, since the charm penguin amplitude is 
fit to data anyway. Some weak annihilation amplitudes have been 
calculated with light-cone QCD sum rules \cite{Khodjamirian:2005wn}; 
the result is compatible with the BBNS parameterization.

It is not difficult to write down the power-suppressed 
operators in SCET~\cite{Feldmann:2004mg}. The problem is that the 
factorization formula involves convolutions, which usually turn out 
to be divergent at the endpoints, making the result meaningless. 
The inadequacy of SCET in addressing this well-known problem 
in hard-exclusive scattering was pointed out in various forms in 
\cite{BF03,Becher:2003kh}, but no solution was offered. In the 
recent paper~\cite{Manohar:2006nz} it is proposed that endpoint 
divergences can be eliminated by a new type of factorization 
(``zero-bin''). This would be a breakthrough; however, I do not 
see how ``zero-bin'' factorization could possibly be correct, since 
it cuts off the endpoint contributions without defining the appropriate 
non-perturbative objects that would represent the endpoint region. 
Thus, the new factorization-scale dependence is not consistently 
canceled.

To explain this I compare the treatment of a certain weak 
annihilation diagram in ``zero-bin'' 
factorization~\cite{Arnesen:2006vb} with the BBNS parameterization 
\cite{BN2003,BBNS2001}. In the first method, the divergent 
integral $\alpha_s\int_0^1 dx \,\phi_{M_2}(x)/\bar x^2$ is 
interpreted as 
\begin{equation}
-\phi^\prime_{M_2}(1)\cdot \alpha_s\ln\frac{m_B}{\mu_-} + F,
\label{ann1}
\end{equation}
in the second as 
\begin{equation}
-\phi^\prime_{M_2}(1)\,\Big(1+\varrho_A e^{i\varphi_A}\Big)
\cdot \alpha_s\ln\frac{m_B}{\Lambda} + F,
\label{ann2}
\end{equation}
where
\begin{equation}
F\equiv \alpha_s \int_0^1 dx\,\frac{\phi_{M_2}(x)+
\bar x \phi_{M_2}^\prime(1)}
{\bar x^2}
\end{equation}
is a finite, subtracted integral. In \cite{Arnesen:2006vb} 
$\mu_-$ is taken to be of order $m_b$, thus the first term 
in (\ref{ann1}) is of order $\alpha_s$ and perturbative. The 
endpoint contribution is effectively set to zero, but the 
dependence on the arbitrary factorization scale 
$\mu_-$ is not canceled. A candidate non-perturbative 
parameter for the endpoint contribution could be $\phi^\prime_{M_2}(1)$, 
but this object is not defined in SCET, so a field-theoretical 
definition of the method is missing. The second expression 
(\ref{ann2}) looks similar, but now there is a large endpoint 
logarithm, and $\alpha_s \ln m_B/\Lambda$ is  of order 1. The 
endpoint contribution is considered to be non-perturbative, and is 
parameterized by the complex quantity $\varrho_A e^{i\varphi_A}$. 
It is again the absence of a field-theoretical definition of this 
quantity that makes the BBNS parameterization a 
phenomenological model. Expression 
(\ref{ann2}) is clearly a more conservative 
treatment of the problem than (\ref{ann1}).

It is evident that in the absence of a field-theoretical definition of 
the zero-bin subtraction method, the statement that ``annihilation 
is real and calculable'' is wishful thinking (I share the wish.); 
it also contradicts the QCD sum rule calculation \cite{Khodjamirian:2005wn}. 
My strong criticism (prompted by strong claims) is not to mean 
that the problem of endpoint factorization is not important. To 
the contrary, its solution is prerequisite to further progress 
in SCET.

\section{PHENOMENOLOGY (OMITTED)}

\noindent 
There is not enough space to discuss the factorization calculations 
of branching fractions and CP asymmetries and the comparison 
with data. I focus on the calculation of the CP-violating 
$S$ parameters and the determination of $\gamma$ in the following 
section. A very brief summary of the other topics discussed in the 
talk reads:

\begin{itemize}
\item The global comparison of all $B\to PP,PV$ data with 
scenario S4 of \cite{BN2003} remains impressively good, including 
CP asymmetries, but there are persistent exceptions. The same is 
true for the PQCD \cite{Li:2005kt,Li:2006jv} and 
BPRS \cite{Williamson:2006hb,Bauer:2005kd} approaches.
\item An enhancement of the electroweak penguin amplitude 
to explain the $\pi K$ system is no longer compelling. 
The difference between the CP asymmetries 
in $\pi^0 K^\pm$ and $\pi^\mp K^\pm$ seems to require an enhancement 
of the colour-suppressed tree amplitude, which cannot be 
explained by factorization.
\item There exist interesting effects \cite{Kagan:2004uw,Beneke:2005we} 
in $B\to VV$ decays, which 
motivate further polarization studies. See \cite{Beneke:2006hg} 
for a comprehensive analysis of these decays. 
\end{itemize}

\vskip-0.7cm
\begin{figure}[h]
\hspace*{-1.5cm}
\includegraphics[width=8cm]{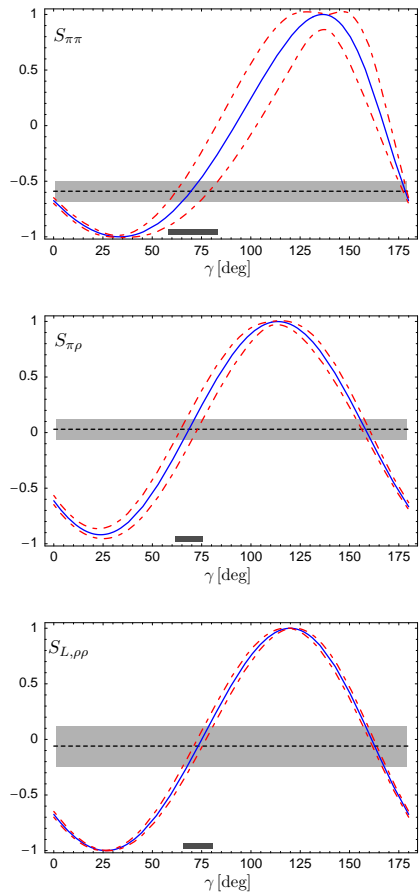}
\vskip-0.7cm
\caption{CKM phase $\gamma$ from $S_f$ with  
$f=(\pi\pi,\pi\rho,[\rho\rho]_L)$.}
\label{fig:S}
\vskip-0.7cm
\end{figure}

\section{\boldmath DETERMINATION OF $\gamma$ FROM $S_f$}

\noindent
The time-dependent CP asymmetries $S_f$ in tree-dominated 
$\Delta D=1$ decays are particularly suited \cite{BN2003,BBNS2001}
to determine the CKM
phase $\gamma$ (or $\alpha$; I assume that $\beta$ is determined 
experimentally) in the framework of QCD factorization, 
since hadronic uncertainty enters only in the 
penguin correction; the dependence on strong phases is reduced, 
because it arises only through $\cos\delta$; the sensitivity to $\gamma$ 
is maximal near $\gamma \sim 70^\circ$.

Figure~\ref{fig:S} shows that for $f=(\pi\pi,\pi\rho,[\rho\rho]_L)$ and 
measurements $S_f=(-0.59\pm 0.09,0.03\pm 0.09, 
-0.06\pm 0.18)$ (HFAG averages), 
one obtains (ignoring a second solution that 
does not lead to consistent results) $\gamma=(70^{+13}_{-10}, 
69 \pm 7, 73\pm 8)^\circ$. The three determinations are 
in agreement with each other, resulting in the 
average $\gamma = (71 \pm 5)^\circ$. 
See \cite{BN2003,Beneke:2006hg} for details.
 
\section{CONCLUSION}

\noindent
The subject of hadronic decays has been and still is a very fertile ground 
for developing new theoretical concepts in heavy flavour physics. 
A lot has been learned about hadronic dynamics. Moreover, 
$\gamma$ is by now known rather well from charmless 
decays. There should be some way to include this information 
in the CKM fits.

\end{document}